\documentclass[aps,prd,preprint,groupedaddress,preprintnumbers,longbibliography,nofootinbib]{revtex4-1}

\newcommand{\nn}{\nonumber}
\newcommand{\lb}{\left\lbrace}
\newcommand{\rb}{\right\rbrace}
\newcommand{\Op}{\mathcal{O}}
\newcommand{\op}[1]{\left[ #1 \right]}

\newcommand{\vvev}[1]{\left\langle\kern-0.3em\left\langle #1
    \right\rangle\kern-0.3em\right\rangle}

\usepackage{amsmath}
\usepackage{graphicx}
\usepackage{amsfonts}

\begin{document}


\preprint{KOBE-TH-17-05}

\title{On the geometry of the theory space in the ERG formalism}


\author{C.~Pagani}
\email[]{capagani@uni-mainz.de}
\affiliation{Institute f\"{u}r Physik (WA THEP)
  Johannes-Gutenberg-Universit\"{a}t\\ Staudingerweg 7, 55099 Mainz,
  Germany}
\author{H.~Sonoda}\email[]{hsonoda@kobe-u.ac.jp}\affiliation{Physics
  Department, Kobe University, Kobe 657-8501, Japan}


\date{\today}

\begin{abstract}
  We consider the theory space as a manifold whose coordinates are
  given by the couplings appearing in the Wilson action.  We discuss
  how to introduce connections on this theory space.  A particularly
  intriguing connection can be defined directly from the solution of the exact renormalization
  group (ERG) equation.  We advocate a geometric viewpoint that lets
  us define straightforwardly physically relevant quantities invariant
  under the changes of a renormalization scheme.
\end{abstract}

\pacs{}

\maketitle


\section{Introduction} \label{sec:intro}

The theory space is a key ingredient of our modern understanding of
quantum and statistical field theory.  On very general grounds, one
may define the theory space as the set of theories that are identified
by the following common features: dimensionality, field content, and
symmetries.  The renormalization group (RG) brings further qualitative
and quantitative information through the notion of relevant,
irrelevant, and marginal directions.  Indeed, the study of the RG flow
of the couplings allows us to define the continuum limit of quantum
field theories and to derive the scaling properties of the operators
by studying the linearized RG flow around a fixed point
\cite{Wilson:1973jj}.

In this work we study the possibility of considering the theory space
as a manifold with geometric structures. In particular, we will show
that it is possible to define connections on the theory space.  The
introduction of a connection is an important step as it allows us to
study in a general way both local and global quantities defined over
the theory space.  We will pay particular attention to a connection
stemming directly and nonperturbatively from the exact
renormalization group (ERG) equation.  By means of a connection, it is
then straightforward to construct the quantities that are invariant
under the changes of coordinates.  A coordinate change can be identified as
a change of schemes (choice of a cutoff function in the ERG framework).
Scheme independence is important since physical observables such as
critical exponents are scheme independent.

In different forms, a geometric viewpoint of the theory space has already been
invoked in the past.  In \cite{Lassig:1989tc}, the RG flow is
identified as a one-parameter group of diffeomorphism generated by the
beta functions as a vector field.  A connection was also identified in
the formulation of renormalization in coordinate space by requiring
covariant transformation properties of the correlation functions
\cite{Sonoda:1993dh,Dolan:1994pq}.
Apart from the linearized behavior around the fixed point, little
effort has been made to investigate seriously the information encoded
in the RG flow beyond critical exponents.  More recently, however, the
transformation properties of RG flows at the second order around a
fixed point have been considered in order to make contact with the
operator product expansion (OPE)
\cite{Lizana:2017sjz,Codello:2017hhh}.  We will comment also on the
relation between our result and the OPE.

The paper is organized as follows.  In
Sec.~\ref{sec:theory-sp-as-manifold} we introduce the theory space as
a manifold and explain its basic features.  In
Sec.~\ref{sec:connection-from-ERG} we consider the ERG equation and
show that its solution implies the existence of a connection and
define its curvature.  In Sec.~\ref{sec:normal-coord-exp} we consider
a covariant expansion of the RG flow and comment on its possible
applications.  In Sec.~\ref{sec:infinite-dim-theory-space} we
generalize our consideration to the full (infinite-dimensional) theory
space.  We summarize our findings in Sec.~\ref{sec:conclusions}.

\section{The theory space as a manifold} \label{sec:theory-sp-as-manifold}

In the Wilsonian renormalization program, one is instructed to write
in the action all possible terms compatible with the symmetries and
the field content of the theory.  Generally, this implies that one has
to consider infinitely many terms in the Wilson action, and
consequently introduce infinitely many couplings.  Therefore, the
theory space is, generally speaking, infinite-dimensional.  However,
if we consider only theories that are defined in the continuum limit,
the actual dimension of the space spanned by the theory is $N$, the
number of relevant directions associated to the fixed point.  In this
work we will mainly consider this latter setting and take a field
theory whose continuum limit is well defined. This permits us to work
with a finite-dimensional manifold.  Some considerations regarding the
infinite-dimensional theory space will be given in
Sec.~\ref{sec:infinite-dim-theory-space}.

Let $g^i\, (i=1,\cdots,N)$ be the $N$ coupling constants parametrizing
the theory.\footnote{ 
Throughout this work the couplings $g^i$ are
taken to be dimensionless as all dimensionful quantities have been
rescaled in units of the cutoff.}  
We view the couplings $g^i$ as coordinates of the theory space and view the latter as a manifold.  
A change of scheme, or cutoff function in the ERG case, results in a possibly very
complicated redefinition of the couplings: $g^\prime{}^i=g^\prime{}^i
\left(g \right)$.  We view such a redefinition as a change of
coordinates on the theory space.
Note that schemes like minimal subtraction are not included straightforwardly 
in the functional RG equations,
although it is known how to retain the former's quantities from the latter,
see \cite{Codello:2013bra} and references therein.
Physical quantities should not depend on the RG scheme employed.
Hence, in the ERG framework, physical quantities should be independent
from the chosen cutoff function, or, equivalently, from the specific coordinates employed.

The RG flow is expressed by the beta functions, which constitute a
vector field over the theory space.  More precisely, a RG trajectory
is described by the beta functions
\begin{equation}
\beta^i = \frac{dg^i}{dt} \quad (i=1,\cdots,N)
\end{equation}
that enjoy the transformation properties of a vector under a coordinate
change.  (We define the ``RG-time'' $t$ by $t\equiv -\log
\frac{\Lambda}{\mu}$, where $\Lambda$ is the cutoff scale introduced
in Sec.~\ref{sec:connection-from-ERG}.)

As we already said, physical quantities must be independent of the RG scheme used
to compute them.  Translated into a geometric language, this means
that physical quantities must be invariant under any change of
coordinates.  An example of such a coordinate invariant quantity is
the critical exponents.  Let us consider
\begin{eqnarray}
  \frac{\partial\beta^{i}}{\partial g^{j}}	&=&
  \frac{\partial}{\partial g^{j}} \sum_{k=1}^N \left(\frac{\partial
      g^{i}}{\partial g^{\prime k}}\beta^{\prime k}\right) 
  \nonumber \\
  &=& \sum_{k,l=1}^N \left(\frac{\partial g^{\prime l}}{\partial
      g^{j}}\frac{\partial^{2}g^{i}}{\partial g^{\prime l}\partial
      g^{\prime k}}\beta^{\prime k}+
    \frac{\partial g^{\prime l}}{\partial
      g^{j}}\frac{\partial\beta^{\prime k}}{\partial g^{\prime
        l}}\frac{\partial g^{i}}{\partial g^{\prime k}}\right)\,. 
\label{eq:trasf-der-beta}
\end{eqnarray}
It is clear that at a fixed point $g^*$ the first term in
(\ref{eq:trasf-der-beta}) vanishes. The critical exponents are defined
as the eigenvalues of the matrix $\partial_{j}\beta^{i}$ at the fixed
point. Since the eigenvalues are independent of the basis used to compute
them, we see that the matrices $\partial_{j}\beta^{i}$ and
$\partial_{j}^{\prime}\beta^{\prime i}$ possess the same spectrum, and
hence yield the same critical exponents.  For later purposes, let us
denote the eigendecomposition of the linearized RG flow at the fixed
point as follows:
\begin{equation}
\frac{\partial\beta^{i}}{\partial g^{j}} \Bigr|_{g=g^*}= \sum_{m,n=1}^N
A_{m}^{i}Y_{n}^{m}\left(A^{-1}\right)_{j}^{n}
\,, \label{eq:eigen-decomp-linearRG} 
\end{equation}
where $Y$ is the eigenvalue matrix, and $A$ is the eigenvector matrix.
It is straightforward to check that $A_{j}^{i}= \sum_{k=1}^N \frac{\partial
  g^{i}}{\partial g^{\prime k}}A_{j}^{\prime k}$.

We note that the coordinate independence of the critical exponents
relies crucially on the vanishing of the inhomogeneous term in
(\ref{eq:trasf-der-beta}) at the fixed point, so that 
the matrix of the linearized RG flow transforms covariantly under a coordinate transformation at the fixed point.  
It is clear, however, that no such simplification occurs when taking further
derivatives of the beta function.  To obviate such difficulties,
instead of employing partial derivatives, it is natural to employ
covariant derivatives that allow us to write down covariant quantities
directly.  It is the purpose of this work to show that such a
geometric structure, namely a connection on the tangent space, can
naturally be introduced from the ERG flow equation.

\section{A connection from the ERGE} \label{sec:connection-from-ERG}

Let $S [\phi]$ be a bare action with an ultraviolet (UV) cutoff
incorporated.  Following \cite{Morris:1993qb}, we introduce
$W_\Lambda [J]$, the generating functional of connected Green
functions with an infrared (IR) cutoff $\Lambda$, by
\begin{equation}
e^{W_\Lambda \left[ J\right]} \equiv \int {\cal D} \phi \, e^{-S\left[
    \phi \right]-\Delta S_\Lambda + \int d^d x\, J \phi} \,, \label{eq:def-Wk} 
\end{equation}
where 
\begin{equation*}
\Delta S_\Lambda =\frac{1}{2} \int d^d x\, \phi (x) R_\Lambda
\left(-\partial^2 \right) \phi (x)
\end{equation*}
is an IR regulator.  The kernel $R_\Lambda \left(-\partial^2 \right)$
suppresses the integration over the modes with momenta lower than the
scale $\Lambda$ in (\ref{eq:def-Wk}). If we denote the Fourier
transform of $R_\Lambda$ by the same symbol $R_\Lambda (p)$, it
approaches a positive constant of order $\Lambda^2$ as $p^2 \to 0$,
and vanishes at large momentum.

The $\Lambda$-dependence of $W_\Lambda$, derived in
\cite{Morris:1993qb}, is given by
\begin{equation}
- \Lambda \frac{\partial W_\Lambda [J]}{\partial \Lambda} = \int_p \Lambda
\frac{\partial R_\Lambda (p)}{\partial \Lambda} \frac{1}{2} \lb
\frac{\delta W_\Lambda [J]}{\delta J (-p)} \frac{\delta W_\Lambda
  [J]}{\delta J(p)} + \frac{\delta^2 W_\Lambda [J]}{\delta J(-p)
  \delta J(p)}\rb\,.
\end{equation}
Here, we wish to consider instead a generalized equation with a
positive anomalous dimension $\eta/2$ for the scalar field
\cite{Igarashi:2016qdr}:
\begin{eqnarray}
- \Lambda \frac{\partial W_\Lambda [J]}{\partial \Lambda} &=& 
\frac{\eta}{2} \int_p J(p) \frac{\delta W_\Lambda [J]}{\delta J(p)}\nn\\
&& + \int_p \left(\Lambda
\frac{\partial}{\partial \Lambda} - \eta \right)
R_\Lambda (p) \cdot \frac{1}{2} \lb
\frac{\delta W_\Lambda [J]}{\delta J (-p)} \frac{\delta W_\Lambda
  [J]}{\delta J(p)} + \frac{\delta^2 W_\Lambda [J]}{\delta J(-p)
  \delta J(p)}\rb\,.\label{WL-dL}
\end{eqnarray}
In the dimensionful convention adopted here, the $N$ parameters of the
theory, say $G^i\,(i=1,\cdots,N)$, do not run as $\Lambda$ changes.
To obtain the running parameters of
Sec.~\ref{sec:theory-sp-as-manifold}, we introduce $\bar{g}^i (t;
G)\,(i=1,\cdots,N)$ as the solution of
\begin{equation}
\frac{\partial}{\partial t} \bar{g}^i (t;G) = \beta^i
\left(\bar{g}\right)\,,
\end{equation}
satisfying the initial condition
\begin{equation}
\bar{g}^i (0;G) = G^i\,.
\end{equation}
We then define
\begin{equation}
g^i \equiv \bar{g}^i \left( - \ln \frac{\Lambda}{\mu}; G \right)\,,
\label{g-via-G}
\end{equation}
where $\mu$ is a reference scale, such that
\begin{equation}
\lim_{\Lambda \to \infty} g^i = g^i_*\,,
\end{equation}
where $g_*$ denotes the fixed point.
These $g$'s are the parameters discussed in
Sec.~\ref{sec:theory-sp-as-manifold}, and they parametrize the theory
in the dimensionless convention.

To switch to the dimensionless convention we divide all physical
quantities by appropriate powers of $\Lambda$ to make them
dimensionless.  We define
\begin{equation}
\bar{J} (p) \equiv \Lambda^{\frac{d-2}{2}} J (p \Lambda)\label{Jbar-J}
\end{equation}
which is a dimensionless field with dimensionless momentum.  We then
define 
\begin{equation}
  W (g) [\bar{J}] \equiv W_\Lambda (G) [J]\,,\label{W-WL}
\end{equation}
where $g$'s are related to $G$'s via (\ref{g-via-G}).  All the
$\Lambda$-dependence of the original functional has been incorporated
into $g$'s and $\bar{J}$.  
We wish to emphasize that we consider only theories in the continuum limit.  
The Wilson action and the functional $W$ have an infinite number of terms, 
but they are related so that these functionals depend only on a finite number of couplings.  
In Appendix \ref{app:cardy-formula}, we give an explicit but perturbative construction of a continuum limit.  
The continuum limit in the ERG framework has been discussed in detail in Ref.~\cite{Rosten:2010vm}.

For fixed $G$'s, we have
\begin{equation}
- \Lambda \frac{\partial}{\partial \Lambda} g^i \Big|_{G} = \beta^i
(g)\,,
\end{equation}
and for fixed $J$, (\ref{Jbar-J}) gives
\begin{equation}
- \Lambda \frac{\partial}{\partial \Lambda} \bar{J} (p)\Big|_J
= \left(\frac{d-2}{2} + p \cdot \partial \right) \bar{J} (p)\,.
\end{equation}
Thus, we obtain
\begin{equation}
- \Lambda \frac{\partial}{\partial \Lambda} W_\Lambda (G) [J]
= \sum_{i=1}^N \beta^i (g) \frac{\partial}{\partial g^i} W(g)
[\bar{J}] + \int_p \left(\frac{d-2}{2}+p \cdot \partial \right) \bar{J} (p)
\frac{\delta}{\delta \bar{J} (p)} W(g) [\bar{J}]\,.
\end{equation}
Hence, (\ref{WL-dL}) implies that $W(g) [\bar{J}]$ obeys the ERG
differential equation 
\begin{eqnarray}
&&\sum_{i=1}^N \beta^i (g) \frac{\partial}{\partial g^i} W(g) [\bar{J}]
= \int_p \left(\frac{d-2+\eta}{2} + p \cdot \partial \right) \bar{J}
(p)
\cdot \frac{\delta W(g)[\bar{J}]}{\delta \bar{J}(p)}\nn\\
&&\quad + \int_p \left( 2 - \eta - p \cdot \partial \right) R (p)\,
\frac{1}{2} \lb \frac{\delta W(g)}{\delta \bar{J} (p)} \frac{\delta
  W(g)}{\delta \bar{J} (-p)} + \frac{\delta^2 W(g)}{\delta \bar{J} (p)
  \delta \bar{J} (-p)} \rb\,,\label{eq:ERG-W}
\end{eqnarray}
where $R(p)$ is related to $R_\Lambda (p)$ of
Sec.~\ref{sec:theory-sp-as-manifold} by
\begin{equation}
R_\Lambda (p) = \Lambda^2 R(p/\Lambda)\,.
\end{equation}
From now on we work only in the dimensionless convention, and we omit the
bar above $J$.

For our purposes, it is useful to think of $W$ as a function of the
couplings, $W= W \left(g\right)$, which is a scalar on the theory
space, $W \left(g\right) = W^\prime \left(g^\prime\right)$.  By taking
a derivative with respect to $g^i$, we obtain
a zero momentum operator
\begin{equation}
\Op_i \equiv \frac{\partial W(g)}{\partial g^i}
\label{eq:def-zero-mom-oper}
\end{equation}
that has covariant transformation properties:
\begin{equation}
\Op_i = \frac{\partial g^{\prime j}}{\partial g^i}\, \Op'_j \,,
\end{equation}
where we have adopted the Einstein convention for repeated indices.

In full analogy we can define the products of the operators ${\cal O}_i$
as follows
\begin{eqnarray}
  \left[{\cal O}_{i_{1}}\cdots{\cal O}_{i_{n}}\right]	\equiv
  e^{-W\left(g\right)}\frac{\partial}{\partial
    g^{i_{1}}}\cdots\frac{\partial}{\partial
    g^{i_{n}}}e^{W\left(g\right)}\,. \label{eq:def-prod-comp-oper} 
\end{eqnarray}
For the case of $\left[{\cal O}_{i_{1}}{\cal O}_{i_{2}}\right]$ we have
\begin{eqnarray}
\left[{\cal O}_{i_{1}}{\cal O}_{i_{2}}\right]	\equiv	\frac{\partial
  W}{\partial g^{i_{1}}}\frac{\partial W}{\partial
  g^{i_{2}}}+\frac{\partial^{2}W}{\partial g^{i_{1}}\partial
  g^{i_{2}}}\,. \label{eq:def-prod-two-comp-oper} 
\end{eqnarray}
Clearly $\left[{\cal O}_{i_{1}}{\cal O}_{i_{2}}\right]$ is not a
covariant quantity.  This is because the 
``connected term''
\begin{equation}
{\cal P}_{ij} \equiv \frac{\partial^{2}W}{\partial g^{i}\partial
  g^{j}} \label{eq:def-conn-part-P} 
\end{equation} 
is not covariant.
Furthermore, $\left[{\cal O}_{i_{1}}{\cal O}_{i_{2}}\right]$ is related to the product of two (zero momentum)
operators, and ${\cal P}_{ij}$ is related to the short distance singularities of this product.
Thus, one expects ${\cal P}_{ij}$ to be related to the OPE's singularities. The precise relation
is hindered by the the fact that we are considering zero momentum operators (i.e.~operators integrated over space).
(A detailed discussion regarding $\left[{\cal O}_{i_{1}}{\cal O}_{i_{2}}\right]$ and ${\cal P}_{ij}$
in the general case of momentum-dependent operators can be found in \cite{Pagani:2017tdr}.)

Now we consider the flow equation for the operators ${\cal O}_i$ and
their products.  The flow of the operator ${\cal O}_i$ can be directly
obtained from (\ref{eq:ERG-W}) by taking a derivative with respect to
$g^i$:
\begin{eqnarray}
\frac{\partial\beta^{k}}{\partial g^{i}}{\cal
  O}_{k}+\left(\beta\cdot\frac{\partial}{\partial g}\right){\cal
  O}_{i} =
{\cal D} {\cal O}_{i} \,, \label{eq:flow-eq-comp-oper-Oj}
\end{eqnarray}
(please recall the Einstein convention for the repeated $k$) where we
define
\begin{eqnarray}
\mathcal{D} &\equiv& \int_p \Bigg[ \left( \frac{d-2+\eta}{2} + p
  \cdot \partial_p \right) J (p) \cdot \frac{\delta}{\delta J(p)}\nn\\
&&\quad + \left(2-\eta-p \cdot \partial \right) R(p) \cdot \lb
\frac{\delta W(g)}{\delta J(-p)} \frac{\delta}{\delta J(p)} +
\frac{1}{2} \frac{\delta^2}{\delta J(p) \delta J(-p)}\rb
\Bigg]\,.\label{eq:def-calD} 
\end{eqnarray}
In deriving (\ref{eq:flow-eq-comp-oper-Oj}) we assume that the
anomalous dimension $\eta$ is independent of $g$'s. This is actually
true only near the fixed point.  The extension to a
$g$-dependent anomalous dimension is given in Appendix
\ref{app-anomalous-dim}.

By taking a further derivative of the flow equation (\ref{eq:ERG-W})
with respect to $g^j$, we deduce the flow equation for ${\cal
  P}_{ij}$.  This can be written as:
\begin{eqnarray}
  &&\frac{\partial^{2}\beta^{k}}{\partial g^{i}\partial g^{j}} {\cal
    O}_{k}+\frac{\partial\beta^{k}}{\partial g^{j}}{\cal
    P}_{ki}+\frac{\partial\beta^{k}}{\partial g^{i}}{\cal P}_{kj}
  +\left(\beta^{k}\frac{\partial}{\partial g^{k}} - {\cal D}
  \right){\cal P}_{ij} 
  = \nonumber  \\  && \qquad \qquad \qquad \qquad \qquad \qquad 
  \int_{p}\left(\left(2-\eta\right)R\left(p^{2}\right)
 -p \cdot \partial_p R\left(p^{2}\right)\right)\frac{\delta 
    {\cal O}_i}{\delta J\left(p\right)}\frac{\delta {\cal O}_j}{\delta 
    J\left(-p\right)}\,. \label{eq:flow-eq-Pij-1}  
\end{eqnarray}
It is interesting to observe that the RHS of (\ref{eq:flow-eq-Pij-1})
is covariant since it is determined by the product of the covariant
operators ${\cal O}_i$ and ${\cal O}_j$.  It follows also that the LHS
of (\ref{eq:flow-eq-Pij-1}) must be covariant, too.

In order to investigate the covariance of the LHS of
(\ref{eq:flow-eq-Pij-1}), let us consider the transformation
properties of ${\cal P}_{ij}$:
\begin{eqnarray}
{{\cal P}^\prime}_{ij} = \frac{\partial g^{k}}{\partial {g}^{\prime
    i}}\frac{\partial g^{l}}{\partial {g}^{\prime j}}{\cal
  P}_{kl}+\frac{\partial^{2}g^{k}}{\partial {g}^{\prime i}\partial
  {g}^{\prime j}} {\cal O}_{k} \,. \label{eq:transf-prop-Pij} 
\end{eqnarray}
${{\cal P}}_{ij} $ is not covariant. Hence, the product $\left[ {\cal
    O}_i {\cal O}_j \right]$ is not covariant as was already pointed
out.  Now we expand ${{\cal P}}_{ij} $ in terms of a basis of
composite operators:
\begin{eqnarray}
{\cal P}_{ij} =
\sum_{k=1}^{N}\Gamma_{i\;j}^{\,\,k}{\cal
  O}_{k}+\sum_{a=N+1}^{\infty}\Gamma_{i\;j}^{\,\,a}{\cal
  O}_{a}\,, \label{eq:def-expansion-Pij} 
\end{eqnarray}
where the operators ${\cal O}_{k}$ with $k\in \left[1,N\right]$ are
the relevant operators conjugate to the couplings $g^k$, whereas the
operators ${\cal O}_{a}$ with $a\in \left[N+1,\infty\right)$ are
irrelevant operators.  By inserting the expansion
(\ref{eq:def-expansion-Pij}) into (\ref{eq:transf-prop-Pij}), we
deduce the transformation properties of the terms appearing in
(\ref{eq:def-expansion-Pij}).  More precisely, we find that
\begin{eqnarray}
{\Gamma}^\prime{}_i{}^k{}_j =
\frac{\partial {g}^{\prime k}}{\partial g^{n}}\frac{\partial
  g^{l}}{\partial {g}^{\prime i}}\frac{\partial g^{m}}{\partial
  {g}^{\prime j}}\Gamma_{l\;m}^{\,\,n}+\frac{\partial {g}^{\prime
    k}}{\partial g^{l}}\frac{\partial^{2}g^{l}}{\partial {g}^{\prime
    i}\partial {g}^{\prime j}}\,, 
\label{eq:conn-from-Pij-transf}
\end{eqnarray}
for $\left(i,j,k \right) \in \left[1,N \right]$ so that
${\Gamma}_i{}^k{}_j$ transforms as a connection in the theory space.
Moreover, we deduce that the second term in
(\ref{eq:def-expansion-Pij}) transforms as a tensor:
\begin{eqnarray}
\sum_{a=N+1}^{\infty} {\Gamma}^\prime{}_i{}^a{}_j  {\cal O}^\prime_{a} =
\frac{\partial g^{k}}{\partial {g}^{\prime i}}\frac{\partial
  g^{l}}{\partial {g}^{\prime
    j}}\sum_{a=N+1}^{\infty}\Gamma_{k\;l}^{\,\,a} {\cal O}_{a}\,. 
\label{eq:tensor-part-Pij-expan} 
\end{eqnarray}
Equation (\ref{eq:def-expansion-Pij}), together with the
transformation properties (\ref{eq:conn-from-Pij-transf}) and
(\ref{eq:tensor-part-Pij-expan}), is one of the main results of this section.
Indeed, our findings entail that, by solving the flow equation, we can
determine a connection over theory space by considering the expansion
of ${\cal P}_{ij}$ in (\ref{eq:def-expansion-Pij}).  Note also that,
by definition, this connection is torsionless, i.e., symmetric in the
lower indices.

It is now natural to come back to Eq.~(\ref{eq:flow-eq-Pij-1}) and
consider its LHS in view of the expansion (\ref{eq:def-expansion-Pij})
and the new connection.  To do so, we also expand the RHS of
(\ref{eq:flow-eq-Pij-1}):
\begin{eqnarray}
  \int_{p}\left(\left(2-\eta\right)R\left(p^{2}\right)
    -p \cdot \partial_p R\left(p^{2}\right)\right)\frac{\delta 
    {\cal O}_i}{\delta J\left(p\right)}\frac{\delta {\cal O}_j}{\delta
    J\left(-p\right)} =
  d_{ij}^k {\cal O}_k +\cdots \,, \label{eq:expansion-rhs-ddW}
\end{eqnarray}
where the dots are contributions involving only irrelevant composite
operators.  In the following we focus our attention solely on the
relevant operators ${\cal O}_i\,(i=1,\cdots,N)$.

As we have already pointed out, the RHS of (\ref{eq:flow-eq-Pij-1}) is
covariant, and the LHS should be also. 
By inserting the expansions
(\ref{eq:def-expansion-Pij}) and (\ref{eq:expansion-rhs-ddW}) into
(\ref{eq:flow-eq-Pij-1}), we find
\begin{eqnarray}
&& \left[\beta^{l}\frac{\partial}{\partial
    g^{l}}\Gamma_{i\;j}^{\,\,k}-\Gamma_{i\;j}^{\,\,l}\frac{\partial\beta^{k}}{\partial
    g^{l}}+\frac{\partial\beta^{l}}{\partial
    g^{j}}\Gamma_{l\;i}^{\,\,k}+\frac{\partial\beta^{l}}{\partial
    g^{i}}\Gamma_{l\;j}^{\,\,k}+\frac{\partial^{2}\beta^{k}}{\partial
    g^{i}\partial g^{j}}\right]{\cal O}_{k}= d_{ij}^k {\cal O}_k
\,, \label{eq:flow-eq-Pij-2} 
\end{eqnarray}
where we have kept only the terms involving relevant operators in the
expansions (\ref{eq:def-expansion-Pij}) and
(\ref{eq:expansion-rhs-ddW}).  The LHS of (\ref{eq:flow-eq-Pij-2}) can
be rewritten in a geometric fashion and, by selecting the term
proportional to ${\cal O}_k$, we can write
\begin{eqnarray}
  \frac{1}{2}\left(\nabla_{i}\nabla_{j}+\nabla_{j}\nabla_{i}\right)\beta^{k}
-\frac{1}{2}\left(R_{il\,\,\,j}^{\,\,\,\,\,k}+R_{jl\,\,\,i}^{\,\,\,\,\,k}\right)\beta^{l} 
  =  d_{ij}^k \,, \label{eq:flow-eq-Pij-3}
\end{eqnarray}
where the covariant derivatives are defined as usual as
\begin{subequations}
\begin{eqnarray}
\nabla_i \beta^j &\equiv& \partial_i \beta^j + \Gamma_{i\;k}^{\,\,j} \beta^k\,,\\
\nabla_i \nabla_j \beta^k &\equiv& \partial_i \left( \nabla_j \beta^k
\right) - \Gamma_{i\;j}^{\,\,l} \nabla_l \beta^k + \Gamma_{i\;l}^{\,\,k}
\nabla_j \beta^l\,,
\end{eqnarray}
\end{subequations}
and the curvature is defined by
\begin{equation}
  R_{il\,\,\,j}^{\,\,\,\,\,k}\equiv\partial_{i}\Gamma_{l\;j}^{\,\,k}-\partial_{l}\Gamma_{i\;j}^{\,\,k}+\Gamma_{i\;m}^{\,\,k}\Gamma_{l\;j}^{\,\,m}-\Gamma_{l\;m}^{\,\,k}\Gamma_{i\;j}^{\,\,m} 
  \,. \label{eq:def-curvature}  
\end{equation}
Equation (\ref{eq:flow-eq-Pij-3}) is one of the main results of this paper.
It shows that the flow equation for ${\cal P}_{ij}$ can be written in
an inspiring covariant form thanks to the connection defined by
Eq.~(\ref{eq:def-expansion-Pij}).  We also wish to point out that
a relation very similar to our Eq.~(\ref{eq:flow-eq-Pij-3}) was
derived in a non-ERG context in \cite{Dolan:1994pq}.  (See also
\cite{Sonoda:1994xg}.)  More details on the derivation of
Eq.~(\ref{eq:flow-eq-Pij-3}) are given in Appendix \ref{app:irr-ope}.

Let us observe that we have constructed the connection
$\Gamma_{i\;j}^{\,\,k}$ using the generating functional $W$.  However,
it can be checked that the same steps can be repeated both for the
Wilson action \cite{Wilson:1973jj,Polchinski:1983gv} and for the
effective average action (EAA)
\cite{Wetterich:1992yh,Ellwanger:1993mw,Morris:1993qb}.

Before concluding this section, we wish to show explicitly that the
curvature defined in (\ref{eq:def-curvature}) is generally
nontrivial.  To see this, let us first consider
\begin{eqnarray}
  \frac{\partial}{\partial g^k}{\cal P}_{ij}
  &=& \partial_{k}\left(\sum_{l=1}^{N}\Gamma_{i\;j}^{\,\,l}{\cal
      O}_{l}+\sum_{a=N+1}^{\infty}\Gamma_{i\;j}^{\,\,a}{\cal
      O}_{a}\right) \nonumber \\ 
  &=&
  \sum_{l=1}^{N}\left(\partial_{k}\Gamma_{i\;j}^{\,\,l}\, {\cal
      O}_{l}+\sum_{m=1}^{N}\Gamma_{i\;j}^{\,\,l}\Gamma_{k\;l}^{\,\,m}{\cal
      O}_{m}+\sum_{a = N+1}^{\infty}\Gamma_{i\;j}^{\,\,l}\Gamma_{k\;l}^{\,\,a}{\cal
      O}_{a}\right) \label{eq:der_k-Pij} \\  
  & &
  +\left(\sum_{a=N+1}^{\infty}\partial_{k}\Gamma_{i\;j}^{\,\,a}\, {\cal
      O}_{a}+\sum_{a=N+1}^{\infty}\Gamma_{i\;j}^{\,\,a}\partial_{k}{\cal
      O}_{a}\right)\,. \nonumber 
\end{eqnarray}
Moreover, it is convenient to consider the following expansion:
\begin{eqnarray}
\partial_{k}{\cal O}_{a>N}	=
\sum_{j=1}^{N}\Gamma_{i\;a}^{\,\,j}\,{\cal
  O}_{j}+\sum_{b=N+1}^{\infty}\Gamma_{i\;a}^{\,\,b}\,{\cal O}_{b}\,.  
\end{eqnarray}
From the definition of ${\cal P}_{ij}$ we deduce
\begin{eqnarray}
\partial_{i}{\cal P}_{kj}&=&\partial_{k}{\cal P}_{ij}
\,. \label{eq:der_k-Pij-der_i-Pkj} 
\end{eqnarray} 
Inserting (\ref{eq:der_k-Pij}) into (\ref{eq:der_k-Pij-der_i-Pkj}) and
extracting the coefficients of the relevant operator ${\cal O}_l$, we
find
\begin{eqnarray}
  \left(\partial_{i}\Gamma_{k\;j}^{\,\,l} 
    +\sum_{m=1}^{N}\Gamma_{k\;j}^{\,\,m}\Gamma_{i\;m}^{\,\,l}\right)
-\left(\partial_{k}\Gamma_{i\;j}^{\,\,l} 
    +\sum_{m=1}^{N}\Gamma_{i\;j}^{\,\,m}\Gamma_{k\;m}^{\,\,l}\right) 
&=& \sum_{a=N+1}^{\infty} \left(
  \Gamma_{i\;j}^{\,\,a}\Gamma_{k\;a}^{\,\,l}-\Gamma_{k\;j}^{\,\,a}\Gamma_{i\;a}^{\,\,l} \right)\,,  
  \nonumber \\
  && \label{eq:rel-from-der-Pij}
\end{eqnarray}
which implies
\begin{eqnarray}
R_{ik\,\,\,j}^{\,\,\,\,\,l} 
=
\sum_{a=N+1}^{\infty} \left( \Gamma_{i\;j}^{\,\,a}\Gamma_{k\;a}^{\,\,l}-\Gamma_{k\;j}^{\,\,a}\Gamma_{i\;a}^{\,\,l} \right)\,.
\label{eq:rel-curvature-GammaGamma}
\end{eqnarray}
Equation (\ref{eq:rel-curvature-GammaGamma}) implies that the curvature is
generally nonzero because there is no reason that the RHS of
(\ref{eq:rel-curvature-GammaGamma}) should vanish.

\section{A different approach: Riemann normal coordinate expansion of the beta functions} \label{sec:normal-coord-exp}

In this section we develop an approach different from the one
considered in Sec.~\ref{sec:connection-from-ERG}, where the
introduction of the connection is deeply related to the flow equation
and its solution.  Here, we wish to consider solely the theory space
manifold and explore it in a covariant way.  As we have argued in
Sec.~\ref{sec:theory-sp-as-manifold}, this is important in order to
define physical, i.e., scheme-independent, quantities. We have already
considered the example of the critical exponents.  The critical
exponents are calculated by considering linear perturbations around
the fixed point.  Nevertheless, information is contained also in the
higher orders of the perturbation, although obtaining scheme invariant
results is hindered by the use of a non-covariant expansion.
Therefore, the purpose of this section is to introduce a covariant
expansion around a fixed point.

Before discussing the nature of the covariant expansion around the
fixed point, we remark that in order to define such an expansion we
need a connection to start with.  In
Sec.~\ref{sec:connection-from-ERG} we have introduced a connection on
the theory space, but this choice is by no means unique. How can we
construct another connection?  There is no canonically defined tensor
like the metric and we have only the vector field defined by the beta
function $\beta^i$.  Given such a vector, it is straightforward to
check that
\begin{equation}
\Gamma_i{}^k{}_j	\equiv	\frac{\partial
  g^{k}}{\partial\beta^{l}}\frac{\partial\beta^{l}}{\partial
  g^{i}\partial g^{j}} \label{eq:flat-connection} 
\end{equation}
transforms as a connection.  (The connection (\ref{eq:flat-connection})
has been also recently proposed in \cite{Lizana:2017sjz}.)

Let us comment on some features regarding this connection.  First of all,
the connection (\ref{eq:flat-connection}) is well defined only when
$\frac{\partial g^{k}}{\partial\beta^{l}}$ actually is.  For the
connection (\ref{eq:flat-connection}) to be defined then, we need
$\frac{\partial g^{k}}{\partial\beta^{l}}$ to be defined. In turn this
implies that the inverse of the matrix $\partial_{i}\beta^{j}$ must
exist.  This inversion can be made locally provided that
$\det \partial_{i}\beta^{j}\neq0$.  In our case of interest, i.e.~in
the vicinity of a fixed point, requiring $\det \partial_{i}\beta^{j}
\neq0$ is tantamount to having no exactly marginal direction.  If an
exactly marginal direction is present, another connection should be
considered.  Furthermore, the connection (\ref{eq:flat-connection}) is
flat as its curvature vanishes identically.  This is a striking
difference from the connection introduced in
Sec.~\ref{sec:connection-from-ERG}.  We will come back to flat
connections in Sec.~\ref{sec:infinite-dim-theory-space}.

Let us now assume that we have some connection $\Gamma_i{}^k{}_j$ and
discuss how to define a covariant expansion for the RG flow by
employing this connection.  The RG flow, as described by the beta
function vector field, is a covariant quantity.  In order to keep
covariance in an expansion, however, special care must be taken.

Quite generally, we are given a vector, which we will later specify to
be $\beta^i$, and we wish to express this vector at some point of the
manifold via a covariant expansion around a different point, which we
will eventually identify with the fixed point.  This reminds us of the
Riemann normal coordinate expansions: given a tensor at some point $P$
(coordinatized by $g^i$), we can express this latter tensor via a
covariant series expansion defined via tensorial quantities evaluated
at the point $Q$ (coordinatized by $g^i_*$, which eventually will be
identified with the fixed point).
More precisely, such an expansion is found by introducing the Riemann
normal coordinates, which we denote $\xi^i$.  The coordinates $\xi^i$
cover a double role: they are a system of coordinates equivalent to
$g^i$, and represent a vector at the point $Q$ coordinatized by
$g^i_*$. In the $\xi$-coordinate system the point $Q$ is represented
by $\xi^i=0$.  We refer the reader to \cite{petrovbook} for more
details.

Applying the Riemann normal coordinate expansion to the vector
$\beta^i$, we obtain
\begin{equation}
\beta^i \left( g \right) = \beta^i \left( g_* \right) + \xi^j
\nabla_j \beta^i \left( g_* \right) 
+\frac{1}{2} \xi^j \xi^k \nabla_j\nabla_k \beta^i
\left( g_* \right) 
+\frac{1}{6}R_{jk}{}^{i}{}_{l} \beta^{j} \left( g_*
\right) \xi^{k}\xi^{l} +\cdots \,.
\label{eq:covariant-RG-pert}
\end{equation}
Note that in order to write down the expansion
(\ref{eq:covariant-RG-pert}) we need to have a connection that defines
the covariant derivative and the curvature.  The same expression holds
for any connection.

Coming back to physical quantities, it is interesting to consider what
information is contained in the second order expansion of the beta
functions.  Let the couplings $\left\lbrace \check{g}^i \right\rbrace$
be conjugate to scaling operators in coordinate space with scaling
dimensions $\Delta_i = D - y_i$, and denote the OPE coefficients
$c_{jk}{}^i$.  Cardy has shown that the beta functions around the
fixed point can be written as \cite{cardy1996scaling}
\begin{equation}
  \check{\beta}^{i}	=
  y_i \check{g}^{i} - \sum_{j,k}c_{jk}{}^i \,
  \check{g}^{j} \check{g}^{k} + O\left(\check{g}^3 \right)
  \,, \label{eq:cardy-formula} 
\end{equation}
where the couplings have been rescaled by an angular integral factor.
One then deduces that
\begin{equation}
\frac{1}{2}\frac{\partial}{\partial \check{g}^{j}}
\frac{\partial}{\partial \check{g}^{k}}
\check{\beta}^{i}\Bigr|_{\check{g}=0}	=	-c_{jk}{}^i
\,. \label{eq:cardy-formula-der} 
\end{equation}
It is natural to ask whether one can use a relation like
(\ref{eq:cardy-formula-der}) in the ERG context.  In this section we
make the first steps in this direction.  (In Appendix
\ref{app:cardy-formula} we also consider the connection of the ERG
with the results of Wegner for the higher order terms in the expansion
of the functional $W\left(g\right)$.)

As it has also been noted in \cite{Codello:2017hhh}, it is crucial to
discuss the dependence of the OPE coefficients on the RG scheme
employed to compute the running of the couplings.  In order to arrive
at a formula involving the scaling fields conjugate to $\left\lbrace
  \check{g}^i \right\rbrace$, we consider the eigendirections of the
linearized RG flow and identify the relation between the couplings
$\left\lbrace \check{g}^i \right\rbrace$ and $\left\lbrace {g}^i
\right\rbrace$ via the matrix $A^{-1}$ introduced in  Eq.~(\ref{eq:eigen-decomp-linearRG}).

However, if we wish to compute the OPE coefficients via Eq.~(\ref{eq:cardy-formula-der}) in terms of $g^i$-dependent quantities,
we see that we have to consider the second derivative
$\partial_{g^{j}}\partial_{g^{k}} {\beta^{i}}$.  More precisely, one
has to consider the following expression: $c_{jk}{}^i \sim
A^{(-1)}{}^i_l \partial_{g^{m}}\partial_{g^{n}} {\beta^{l}} A^m_j A^n_k$.  
From the transformation properties of $A$ and $\beta$ it is
straightforward to check that the so defined $c_{jk}{}^i$ is invariant
under coordinate transformations up to an additive term due to the
fact that $\partial_{g^{m}}\partial_{g^{n}} {\beta^{l}}$ does not
transform as a tensor (see also \cite{Codello:2017hhh}).

To obviate this fact one may consider the covariant version of $\partial_{g^{j}}\partial_{g^{k}} {\beta^{i}}$, 
where the partial derivatives have been promoted to covariant derivatives: $\nabla_{g^{m}}\nabla_{g^{n}} {\beta^{l}}$.  
It is clear then that the expression $ A^{(-1)}{}^i_l
\nabla_{g^{m}}\nabla_{g^{n}} {\beta^{l}} A^m_j A^n_k$ is invariant
under a change of scheme and thus it is a physical candidate to be
considered.  The purpose of the geometric expansion
(\ref{eq:covariant-RG-pert}) is exactly to probe the vicinity of the
fixed point in a covariant fashion, and it provides a natural
introduction for the covariant expression
$\nabla_{g^{m}}\nabla_{g^{n}} {\beta^{l}}$.  Critical exponents are
found by looking at the linear perturbation around the fixed point,
which corresponds to the first term in (\ref{eq:covariant-RG-pert})
where $\xi$ corresponds to the perturbation.  The second term in
(\ref{eq:covariant-RG-pert}) now contains the information regarding
the second order perturbation around the fixed point in a covariant
manner.

We conclude this section by stressing that the covariant expansion
(\ref{eq:covariant-RG-pert}) can be used in the ERG context to define
further physical quantities besides the critical exponents, such as
the Wilson operator product coefficients.  Nevertheless, employing
different connections selects different quantities, and it is not
straightforward to deduce their meaning.  However, the discussion of
the previous section and its connection with the previous works in the
literature, e.g.~\cite{Sonoda:1994xg}, suggest that OPE coefficients
are found by employing the connection of
Sec.~\ref{sec:connection-from-ERG}.

\section{The infinite-dimensional theory space} \label{sec:infinite-dim-theory-space}

So far we have taken the theory space to be $N$ dimensional, with $N$
being the number of relevant directions.  This is possible solely for
renormalizable trajectories, that is, theories whose continuum limit is
well defined.  However, the ERG framework can be employed to test the
theory space with its fullest content, i.e., taking into account also
the infinitely many irrelevant directions.  The aim of this section is
to discuss how the machinery developed until now is modified when
considering this more general theory space.

In actual applications of the ERG, the need for an ansatz or some
truncation scheme generally requires us to consider a finite-dimensional
approximation of the theory space, which is then parametrized by $n$
couplings with $N$ relevant and $n-N$ irrelevant directions.  For the
purposes of this section, let us consider $n$ fixed and eventually
take the formal limit $n\rightarrow \infty$.

The definition of the connection (\ref{eq:flat-connection}) can be
straightforwardly extended by truncating the theory space to include
the $n-N$ irrelevant directions.  In a typical ERG computation, where
an ansatz $S_\Lambda = \sum_{i=1}^n g^i {\cal O}_i $ is considered, we
have $n$ coordinates and beta functions, and a connection may be
considered.

Let us go back to the framework developed in
Sec.~\ref{sec:connection-from-ERG}, and adapt it to the present $n-$dimensional space.  The expansion (\ref{eq:def-expansion-Pij}) of
${\cal P}_{ij}$ is no longer split in relevant and irrelevant parts,
but we include all the operators in a single sum (possibly truncated,
retaining only $n$ operators).  Extending the range of indices of the
connection is not as innocuous as it may seem.  Indeed, by repeating the
reasoning at the end of Sec.~\ref{sec:connection-from-ERG} stemming
from the relation $\partial_k {\cal P}_{ij} = \partial_i {\cal
  P}_{kj}$ we see that now the curvature identically vanishes.  This
is due to the inclusion of the RHS of (\ref{eq:rel-from-der-Pij}) in the
definition of the curvature.

Is there any obvious reason for this fact?  Let us consider that we
can view the theory space as a space of functionals, i.e., the
Wilsonian actions $S_\Lambda$, and that there is a priori no need for
this space to be flat.  However, if we {\it assume} that such
functionals can be expanded in couplings as $S_\Lambda= \sum_i g^i
{\cal O}_i$, where the ${\cal O}_i$ are independent of $g^i$, we can
check that this space enjoys the properties of a vector space, e.g.,
distributivity $\sum_i g^i {\cal O}_i + \sum_i \tilde{g}^i {\cal
  O}_i=\sum_i \left(g^i+\tilde{g}^i\right) {\cal O}_i$.  Any
$n$-dimensional vector space is isomorphic to $\mathbb{R}^{n}$, which
is a flat space.  Thus, in this sense, it is appealing to consider the
theory space as a flat manifold.

This is a striking difference from the ``continuum theories subspace''
considered in Sec.~\ref{sec:connection-from-ERG}.  However, this is
not a contradiction.  Actually, even if the full theory space were
flat, it would be generally possible to have a curved subspace
expressed in the intrinsic coordinates provided by the relevant
couplings $g^i$ with $i=1,\cdots ,N$.

In the ``continuum theories subspace'' one could possibly consider
non-trivial topological invariants.  For instance, for a subspace of
dimension $N=2p$ one could consider the Euler invariant
\begin{eqnarray}
E_{2p} =
\frac{\left(-1\right)^{p}}{2^{2p}\pi^{p}p!}\int\epsilon_{i_{1}\cdots
  i_{2p}} R^{i_{1}i_{2}}\wedge\cdots\wedge R^{i_{2p-1}i_{2p}} 
\end{eqnarray}
which is defined via the exterior product of $p$ curvature two-forms
$R$ defined in (\ref{eq:def-curvature}).  It is not clear, though, if
the above $E_{2p}$ could be of any practical interest.

\section{Conclusions} \label{sec:conclusions}

In this work we have put forward a geometric viewpoint on the theory
space inspired by the ERG flow equation.  While viewing the theory
space as a manifold, we have introduced further geometric structures.
In particular we have shown it possible to define connections over the
theory space.  The theory space has been, for most of this work,
restricted to the space where the continuum limit of the field theory
is well defined.

Remarkably, we have been able to define explicitly two connections. 
One stems from the expansion of ${\cal P}_{ij}$ in
composite operators ${\cal O}_k$; see
Eqs.~(\ref{eq:def-expansion-Pij}) and (\ref{eq:conn-from-Pij-transf}).
The other exploits the transformation properties of the beta
functions; see Eq.~(\ref{eq:flat-connection}).  In
Sec.~\ref{sec:connection-from-ERG} we have also shown that the ERG
equation associated with the expansion (\ref{eq:def-expansion-Pij})
can be written in a manifestly covariant way.

In Sec.~\ref{sec:normal-coord-exp} we have discussed a different
geometric view on the RG flow.  Namely, we have looked at the RG flow
around the fixed point via a covariant expansion by employing the
Riemann normal coordinates.  Furthermore, we have emphasized that our
geometric framework allows us to possibly define further physical
quantities directly from the RG flow.  In this case, physical
quantities are identified as scheme-independent quantities, such as
the critical exponents.

In Sec.~\ref{sec:infinite-dim-theory-space} we have considered the
full (infinite-dimensional) theory space.  We have noted that the full
theory space is actually flat and that one may view the
``renormalizable theories subspace'' as a curved submanifold embedded
in the full (flat) theory space.

Concluding this paper, we would like to remark that the geometric
understanding of the theory space, introduced here, could be
helpful in defining in a suitable manner further physical quantities,
such as the operator product expansion coefficients, on top of the
critical exponents.
In the future, we hope to be able to come back to the formalism developed in this work
and compute explicitly some of the quantities that we have introduced,
like the connection $\Gamma_i{}^k{}_j$ and the associated curvature, 
in some approximation scheme (e.g.~epsilon or $1/N$ expansion).

\appendix

\section{Inclusion of the anomalous dimension\label{app-anomalous-dim}}

In Sec.~\ref{sec:connection-from-ERG} we derived the geometric
relation (\ref{eq:flow-eq-Pij-3}) while neglecting the coupling dependence
of the anomalous dimension. Here we generalize Eq.~(\ref{eq:flow-eq-Pij-3}) by including such dependence.

The anomalous dimension $\eta = \eta \left(g\right)$ is a scalar under
coordinate transformations.  It follows that a derivative
$\partial_i \eta =\nabla_i \eta$ is a covariant quantity, whereas a
second derivative is not.  By taking a derivative with respect to
$g^j$ of (\ref{eq:ERG-W}) we obtain
\begin{eqnarray}
&& \frac{\partial\beta^{i}}{\partial g^{j}} \,{\cal
   O}_{i}+\left(\beta\cdot\frac{\partial}{\partial g}\right){\cal
   O}_{j} = {\cal D} {\cal O}_{j}
+\int_p \frac{1}{2}\frac{\partial\eta}{\partial g^{j}} J\left(p\right)
   \frac{\delta W}{J\left(p\right)}
  \\ \label{eq:flow-eq-comp-oper-Oj-eta} 
&& \quad \quad \quad \quad \quad \quad
+\frac{1}{2}\int_p \left(-\frac{\partial\eta}{\partial g^{j}}R \left(p^2 \right) \right)
\left[\frac{\delta W}{\delta J\left(p \right)}\frac{\delta W}{\delta
   J\left(-p \right)}+\frac{\delta^{2}W}{\delta J\left(p \right)\delta
   J\left(-p \right)}\right] 
 \,, \nonumber
\end{eqnarray}
which is equivalent to Eq.~(\ref{eq:flow-eq-comp-oper-Oj}) when $\eta$ is a constant.
Equation (\ref{eq:flow-eq-comp-oper-Oj-eta}) can be written in a more geometric fashion as follows:
\begin{eqnarray}
&& \nabla_j \beta^{i} {\cal O}_{i}+ \beta^{i} \nabla_{i} {\cal O}_{j}
=
{\cal D} {\cal O}_{j}
+\nabla_j\eta \int_p \frac{1}{2}   J\left(p\right) \frac{\delta W}{J\left(p\right)}  \nonumber \\ 
&& \quad \quad \quad \quad \quad \quad
-\frac{1}{2} \nabla_j\eta  \int_p R\left(p^2 \right)
\left[\frac{\delta W}{\delta J\left(p \right)}\frac{\delta W}{\delta J\left(-p \right)}+\frac{\delta^{2}W}{\delta J\left(p \right)\delta J\left(-p \right)}\right]
 \,, \nonumber
\end{eqnarray}
where we used the fact that the connection is symmetric.

By differentiating once again with respect to $g^i$ we obtain
\begin{eqnarray}
\beta\cdot\frac{\partial}{\partial g}{\cal P}_{ij}-\frac{\partial\beta^{k}}{\partial g^{j}}{\cal P}_{ki}+\frac{\partial\beta^{k}}{\partial g^{i}}{\cal P}_{kj}+\frac{\partial\beta^{k}}{\partial g^{i}\partial g^{j}}O_{k}
=
\mbox{RHS} \label{eq:flow-Pji-eta}
\end{eqnarray}
where
\begin{eqnarray}
  \mbox{RHS}
  &=&
  {\cal D} P_{ij} 
  +\int_{p}\left(\left(2-\eta\right)R\left(p^{2}\right)-p
    \cdot \partial_p R\left(p^{2}\right)\right)\frac{\delta {\cal
      O}_i}{\delta J\left(p\right)}\frac{\delta {\cal O}_j}{\delta
    J\left(-p\right)} \nonumber \\ 
  &&
  +\frac{1}{2}\frac{\partial\eta}{\partial g^{i}} \int_p
  J\left(p\right) \frac{\delta}{\delta J\left(p\right)} \frac{\partial
    W}{\partial g^{j}} 
  +\frac{1}{2}\frac{\partial\eta}{\partial g^{j}} \int_p
  J\left(p\right) \frac{\delta}{\delta J\left(p\right)} \frac{\partial
    W}{\partial g^{i}}  
+\frac{1}{2}\frac{\partial^2 \eta}{\partial g^{i} \partial g^{j}}
\int_p J\left(p\right) \frac{\delta W}{\delta J\left(p\right)}
\nonumber \\ 
&& -\frac{\partial^2 \eta}{\partial g^{i} \partial g^{j}} \int_p
R\left(p^2\right)\left[\frac{1}{2}\frac{\delta W}{\delta
    J\left(p\right)}\frac{\delta W}{\delta
    J\left(-p\right)}+\frac{1}{2}\frac{\delta^{2}W}{\delta
    J\left(p\right) \delta J\left(-p\right)}\right] \nonumber \\ 
&& -\frac{\partial\eta}{\partial g^{j}} \int_p R\left(p^2\right)
\left[\frac{\delta W}{\delta J\left(-p\right)} \frac{\delta {\cal
      O}_i}{\delta J \left(p\right)}+\frac{1}{2}\frac{\delta^{2} {\cal
      O}_i}{\delta J\left(p\right) \delta J\left(-p\right)} \right]
\nonumber \\ 
&&  -\frac{\partial\eta}{\partial g^{i}} \int_p R\left(p^2\right)
\left[\frac{\delta W}{\delta J\left(-p\right)} \frac{\delta {\cal
      O}_j}{\delta J\left(p\right)}+\frac{1}{2}\frac{\delta^{2} {\cal
      O}_j}{\delta J\left(p\right) \delta J\left(-p\right)} \right]
\nonumber  
\end{eqnarray}
Following the same steps as in Sec.~\ref{sec:connection-from-ERG},
using Eq.~(\ref{eq:flow-eq-comp-oper-Oj-eta}), and dropping terms
coming from irrelevant operators we can rewrite
(\ref{eq:flow-Pji-eta}) as follows
\begin{eqnarray}
&& \left[\frac{1}{2}\left(\nabla_{i}\nabla_{j}+\nabla_{j}\nabla_{i}\right)\beta^{k}-\frac{1}{2}\left(R_{il\,\,\,j}^{\,\,\,\,\,k}+R_{jl\,\,\,i}^{\,\,\,\,\,k}\right)\beta^{l} \right]  {\cal O}_k=
d_{ij}^k {\cal O}_k   \label{eq:geom-flow-ddW-eta} \\
&& \quad \quad
+\frac{1}{2} \nabla_i \eta \int_p J\left(p\right) \frac{\delta}{\delta J\left(p\right)} \frac{\partial W}{\partial g^{j}}
+\frac{1}{2} \nabla_j \eta \int_p J\left(p\right) \frac{\delta}{\delta J\left(p\right)} \frac{\partial W}{\partial g^{i}} 
+\frac{1}{2} \nabla_i \nabla_j \eta \int_p J\left(p\right) \frac{\delta W}{\delta J\left(p\right)} \nonumber \\
&& \quad \quad
- \nabla_i \nabla_j \eta \int_p R\left(p^2\right)\left[\frac{1}{2}\frac{\delta W}{\delta J\left(p\right)}\frac{\delta W}{\delta J\left(-p\right)}+\frac{1}{2}\frac{\delta^{2}W}{\delta J\left(p\right) \delta J\left(-p\right)}\right] \nonumber \\
&& \quad \quad
- \nabla_j \eta \int_p R\left(p^2\right) \left[\frac{\delta W}{\delta J\left(-p\right)} \frac{\delta {\cal O}_i}{\delta J \left(p\right)}+\frac{1}{2}\frac{\delta^{2} {\cal O}_i}{\delta J\left(p\right) \delta J\left(-p\right)} \right] \nonumber \\
&& \quad \quad
  - \nabla_i \eta \int_p R\left(p^2\right) \left[\frac{\delta W}{\delta J\left(-p\right)} \frac{\delta {\cal O}_j}{\delta J\left(p\right)}+\frac{1}{2}\frac{\delta^{2} {\cal O}_j}{\delta J\left(p\right) \delta J\left(-p\right)} \right] \nonumber \,.
\end{eqnarray}
The first line in (\ref{eq:geom-flow-ddW-eta}) corresponds to (\ref{eq:flow-eq-Pij-3}) for 
the case of constant $\eta$.
As in the case of Eq.~(\ref{eq:expansion-rhs-ddW}), 
the $\eta$-dependent lines in (\ref{eq:geom-flow-ddW-eta}) can be expanded in the ${\cal O}_k$ basis,
retaining only the relevant operators.

\section{The role of irrelevant operators in (\ref{eq:flow-eq-Pij-3})} \label{app:irr-ope}

In deriving Eq.~(\ref{eq:flow-eq-Pij-3}) we truncated the
expansion (\ref{eq:def-expansion-Pij}) for ${\cal{P}}_{ij}$ by
retaining only the relevant operators.  One may wonder if any effect
is to be expected from the irrelevant operators, since the RG flow of
irrelevant operators mixes in general with relevant ones.  In this
appendix we discuss this point in detail.

Let us first introduce irrelevant composite operators.  From the
transformation property (\ref{eq:tensor-part-Pij-expan}) we deduce
that an irrelevant operator is a scalar quantity labeled by an index
$a\in \left[N+1,\infty \right)$.  Such index then cannot be traced
back to a coordinate index, rather it can be thought of as an
``internal index''.  For this reason, in this section we shall denote
the composite operators via greek indices $\mu = a\in \left[N+1,\infty
\right)$.  Adopting this notation we can write the coordinate
transformation property (\ref{eq:tensor-part-Pij-expan}) as
\begin{eqnarray}
{\Gamma}^\prime{}_i{}^\mu{}_j  {\cal O}^\prime_{\mu} =
\frac{\partial g^{k}}{\partial {g}^{\prime i}}\frac{\partial
  g^{l}}{\partial {g}^{\prime j}} \Gamma_{k\; l}^{\,\,\mu} {\cal
  O}_{\mu}\,, 
\end{eqnarray}
where the sum over $\mu$ is intended.  An operator ${\cal O}_{\mu}$
transforms as a scalar, and $\Gamma_{i\; j}^{\,\,\mu}$ transforms as a tensor in
the two lower indices.  Furthermore, an operator ${\cal O}_{\mu}$
satisfies the following ERG equation:
\begin{equation}
\left( \beta \cdot \frac{\partial}{\partial g} - {\cal D} \right)
{\cal O}_{\mu} +y_\mu {\cal O}_{\mu}  
= M_\mu{}^i {\cal O}_{i} +M_\mu{}^\nu {\cal
  O}_{\nu}\,, \label{eq:ERG-equation-irr-oper} 
\end{equation}
where we split the mixing into relevant and irrelevant operators in
the RHS.  From the transformation properties of ${\cal O}_{i}$ and
${\cal O}_{\mu}$, we deduce that the matrix $M_\mu{}^i$ transforms as a
vector.  Moreover, at the fixed point, the ERG equation
(\ref{eq:ERG-equation-irr-oper}) reduces to
\begin{equation}
\left(y_\mu - {\cal D} \right) {\cal O}_{\mu} =
0\,, \label{eq:ERG-equation-irr-oper-FP} 
\end{equation}
where $-y_\mu \geq 0$ is the scaling dimension of ${\cal O}_\mu$ in
momentum space.

Employing the notation introduced so far, we can rewrite the expansion
(\ref{eq:def-expansion-Pij}) as follows:
\begin{eqnarray}
{\cal P}_{ij} = \Gamma_{i\;j}^{\,\,k}{\cal O}_{k} +
\Gamma_{i\;j}^{\,\,\mu}{\cal
  O}_{\mu}\,. \label{eq:expansion-Pij-appendix} 
\end{eqnarray}
Then, plugging the expansion (\ref{eq:expansion-Pij-appendix}) into
(\ref{eq:flow-eq-Pij-1}), it is straightforward to check that a new
term appears in (\ref{eq:flow-eq-Pij-3}).  Such a term arises due to the
following contribution:
\begin{eqnarray*}
\left(\beta^{k}\frac{\partial}{\partial g^{k}}-{\cal D}\right) {\cal P}_{ij}	
&\supset &
\Gamma_{i\;j}^{\,\,\mu}\left(\beta^{k}\frac{\partial}{\partial
    g^{k}}-{\cal D}\right){\cal O}_{\mu} \\ 
& = & \Gamma_{i\;j}^{\,\,\mu}\left(-y_{\mu}{\cal
    O}_{\mu}+M_{\mu}^{\,\,\,k}{\cal O}_{k}+M_{\mu}^{\,\,\,\nu}{\cal
    O}_{\nu}\right)\,. 
\end{eqnarray*}
Thus we see that also a term proportional to the relevant operator
${\cal O}_{k}$ is generated and that Eq.~(\ref{eq:flow-eq-Pij-3})
is generalized to
\begin{eqnarray}
\frac{1}{2}\left(\nabla_{i}\nabla_{j}+\nabla_{j}\nabla_{i}\right)\beta^{k}-\frac{1}{2}\left(R_{il\,\,\,j}^{\,\,\,\,\,k}+R_{jl\,\,\,i}^{\,\,\,\,\,k}\right)\beta^{l} 
+\Gamma_{i\;j}^{\,\,\mu} M_{\mu}^{\,\,\,k} 
= d_{ij}^k \,, \label{eq:geom-formula-appendix}
\end{eqnarray}
where the last term on the LHS transforms also as a tensor.
Note that at a fixed point Eq.~(\ref{eq:geom-formula-appendix}) reads
\begin{eqnarray*}
\frac{1}{2}\left(\nabla_{i}\nabla_{j}+\nabla_{j}\nabla_{i}\right)\beta^{k}
\Bigr|_{\rm{FP}} 
= d_{ij}^k \Bigr|_{\rm{FP}} \,,
\end{eqnarray*}
since the last term in (\ref{eq:geom-formula-appendix}) does not
contribute to the fixed point formula.

Now let us discuss in more detail the presence of the term
$\Gamma_{i\;j}^{\,\,\mu} M_{\mu}^{\,\,\,k}$ in
(\ref{eq:geom-formula-appendix}).  In particular, we wish to make two
observations which reveal that $\Gamma_{i\;j}^{\,\,\mu}
M_{\mu}^{\,\,\,k}$ constitutes a subleading contribution to
(\ref{eq:geom-formula-appendix}).

The first observation is based on an explicit estimate of the cutoff
dependence in the dimensionful convention.  A careful analysis, based
on the choice of coordinates found in \cite{Sonoda:1990gp,Sonoda:1991mk},
shows that the contribution due to the irrelevant operators in
(\ref{eq:expansion-Pij-appendix}) is subleading in the large
$\Lambda$ limit.  More precisely, denoting $y_O \equiv d -\Delta_O$,
where $\Delta_O$ is the scaling dimension of an operator
$O\left(x\right)$, the leading contributions scale like
$\Lambda^{y_k-y_i-y_j}$.  For $y_k>y_i+y_j$, this leads to a singular
behavior that can be put in correspondence with the nonintegrable
short distance singularities in the OPE via dimensional analysis
arguments.  The term $\Gamma_{i\;j}^{\,\,\mu} M_{\mu}^{\,\,\,k}$ does
not contribute to the singular behavior and can be dropped in
(\ref{eq:geom-formula-appendix}) when considering nonintegrable short
distance singularities as it scales like
$\Lambda^{\left(y_\mu-y_i-y_j\right)<0}$.  This observation makes
evident a link with some previous works in the literature (see in
particular
\cite{Sonoda:1991mv,Sonoda:1993dh,Sonoda:1994qg,Sonoda:1996ta}), where
the nonintegrable short distance singularities are considered, and a
geometric formula fully analogous to (\ref{eq:flow-eq-Pij-3}) is
derived.

As a second observation, we note that in order to write down
(\ref{eq:geom-formula-appendix}) a certain basis of irrelevant
operators has been selected.  If we limit ourselves to consider
nonintegrable short distance singularities, i.e., scaling dimensions
such that $y_k > y_i +y_j$, then the term $\Gamma_{i\;j}^{\,\,\mu}
M_{\mu}^{\,\,\,k}$ is dismissed.  Hence this truncation has the nice
feature of being independent of the convention chosen for the
irrelevant operators.

\section{Cardy's formula} \label{app:cardy-formula}

Let us consider a generic fixed point with $N$ relevant directions.
Following \cite{Wegner:1972} we construct the Wilson action
perturbatively around the fixed point.  Let us denote the relevant
parameters with scale dimension $y_i > 0$ by $g^i\, (i=1,\cdots,N)$.
The generating functional $W(g)$ with an IR cutoff is determined by
\begin{eqnarray}
&&\sum_{i=1}^N \beta^i (g) \frac{\partial}{\partial g^i} e^{W(g)[J]}
= \int_p \left[ \left( p \cdot \partial_p + \frac{D-2}{2} + \gamma
  \right) J(p) \cdot \frac{\delta}{\delta J(p)} \right.\nn\\
&&\quad\left. + \left( - p \cdot \partial_p + 2 - 2 \gamma \right)
  R(p) \cdot \frac{1}{2} \frac{\delta^2}{\delta J(p) \delta J(-p)}
\right] e^{W(g) [J]}\,.
\end{eqnarray}
Denoting the fixed point functional $W^* = W(g=0)$, we rewrite this in
a form more convenient for perturbative calculations:
\begin{eqnarray}
&&\sum_{i=1}^N \beta^i (g) \frac{\partial}{\partial g^i} e^{W(g)-W^*}
= \int_p \left[ \left( p \cdot \partial_p + \frac{D-2}{2} +
    \gamma \right) J(p) \frac{\delta}{\delta J(p)} \right.\nn\\
&&\left.\quad + \left( - p \cdot \partial_p + 2 - 2 \gamma \right) R(p)
\cdot \left( \frac{\delta W^* [J]}{\delta J(-p)} \frac{\delta}{\delta J(p)}
+ \frac{1}{2} \frac{\delta^2}{\delta J(p) \delta J(-p)} \right)\right]
\, e^{W(g)-W^*}\,.\label{ERGeq-W(g)}
\end{eqnarray}
We assume a constant anomalous dimension $\gamma$ for simplicity.  We
wish to solve this perturbatively by expanding the functional as
\begin{subequations}
\label{perturbative-expansions}
\begin{equation}
W (g) = W^* + \sum_{i=1}^N g_i W_i + \sum_{i,j=1}^N \frac{1}{2} g_i
g_j W_{ij} + \sum_{i,j,k=1}^N \frac{1}{3!} g_i g_j g_k W_{ijk} +
\cdots\,.\label{W-expansion}
\end{equation}
and the beta functions as
\begin{equation}
\beta^i (g) = y_i g^i + \frac{1}{2} \sum_{j,k=1}^N \beta^i_{jk} g^j
g^k + \frac{1}{3!} \sum_{j,k,l=1}^N \beta^i_{jkl} g^j g^k g^l +
\cdots\,.\label{beta-expansion}
\end{equation}
\end{subequations}

We can regard $g^i$ as the coefficient of an external source with zero
momentum.  Hence,
\begin{equation}
\Op_{i_1, \cdots, i_n} =  e^{- W^*} \frac{\partial^n}{\partial
  g^{i_1} \partial g^{i_2} \cdots \partial g^{i_n}}
                           e^{W(g)}\Big|_{g=0}
\end{equation}
is the $n$th order product of composite operators $W_i =
\frac{\partial}{\partial g^i} W(g)\Big|_{g=0}$ with zero momentum.  We
obtain, up to third order,
\begin{subequations}
\begin{eqnarray}
\Op_i &=& W_i\,,\\
\Op_{ij} &=& \op{\Op_i \Op_j} = \Op_i \Op_j + W_{ij}\,,\\
\Op_{ijk} &=& \op{\Op_i \Op_j \Op_k} = \Op_i \Op_j \Op_k + W_{ij}
\Op_k + W_{ik} \Op_j + W_{jk} \Op_i + W_{ijk}\,.
\end{eqnarray}
\end{subequations}
$\Op_{i_1,\cdots, i_n}$ satisfies the ERG equation
\begin{eqnarray}
\left(- \sum_{j=1}^n y_{i_j} + \mathcal{D}\right) \Op_{i_1 \cdots
  i_n}
&=& \sum_{j=1}^N \Bigg[ \sum_{1 \le \alpha < \beta \le n}  \beta^j_{i_\alpha i_\beta}
\Op_{j i_1 \cdots \widehat{i_\alpha} \cdots \widehat{i_\beta} \cdots
  i_n}\nn\\
&& \quad+ \sum_{1 \le \alpha_1 < \alpha_2 < \alpha_3 \le n} 
\beta^j_{i_{\alpha_1} i_{\alpha_2} i_{\alpha_3}} 
\Op_{j i_1
  \cdots \widehat{i_{\alpha_1}} \cdots \widehat{i_{\alpha_2}} \cdots
  \widehat{i_{\alpha_3}} \cdots i_n}\nn\\
&& \quad+ \cdots \nn\\
&& \quad+ \beta_{j, i_1 \cdots i_n} \Op_j\, \Bigg]\,,
\label{ERGeq-On}
\end{eqnarray}
where $\mathcal{D}$ is the functional differential operator defined by
the right-hand side of (\ref{ERGeq-W(g)}).  We have thus shown
that the higher order derivatives of the beta functions give mixing of
the operator products.

We only consider the first two cases: $n=1, 2$.  Taking $n=1$ in
(\ref{ERGeq-On}), we obtain
\begin{equation}
\left( y_i - \mathcal{D} \right) W_i = 0\,,\quad (i=1,\cdots,N)
\end{equation}
implying that $W_i$ is a composite operator of scale dimension $-y_i$.
(This was actually taken for granted.)  Taking $n=2$ in
(\ref{ERGeq-On}), we obtain
\begin{equation}
\left( y_j + y_k - \mathcal{D}\right) W_{jk} = - \sum_{i=1}^N W_i
\beta^i_{jk} + \int_p \left( - p \cdot \partial_p + 2 - 2 \gamma
\right) R(p) \cdot \frac{\delta W_j}{\delta J(p)} \frac{\delta
  W_k}{\delta J(-p)}\,.
\end{equation}
The integral is local, and we can expand
\begin{equation}
 \int_p \left( - p \cdot \partial_p + 2 - 2 \gamma
\right) R(p) \cdot \frac{\delta W_j}{\delta J(p)} \frac{\delta
  W_k}{\delta J(-p)} = \sum_{i=1}^\infty d_{jk}^i \, \Op_i\,,
\end{equation}
where $\Op_i = W_i\, (i=1,\cdots,N)$, and $\Op_{i > N}$ are irrelevant
operators of scale dimension $- y_i \ge 0$.  Hence, we obtain
\begin{equation}
\left( y_j + y_k - \mathcal{D}\right) W_{jk} = \sum_{i=1}^N W_i \left(
  d_{jk}^i - \beta^i_{jk} \right)
+ \sum_{i > N} d_{jk}^i \, \Op_i\,.
\end{equation}

In the absence of degeneracy, i.e., 
\begin{equation}
y_j + y_k \ne y_i
\end{equation}
for any $i,j,k \le N$, we can choose
\begin{equation}
\beta^i_{jk} = 0\label{appendix-linear-beta}
\end{equation}
so that 
\begin{equation}
W_{jk} = \sum_{i=1}^\infty \frac{d_{jk}^i}{y_j+y_k-y_i}\, \Op_i\,.
\end{equation}
Hence, the beta functions are linear up to second order.  This is
expected from the old result of Wegner \cite{Wegner:1972}.  (In the
absence of degeneracy, the parameters can be chosen to satisfy linear
RG equations.)

Alternatively, we can demand $W_{jk}$ be free of
$W_i\,(i=1,\cdots,N)$.  We must then choose
\begin{equation}
\beta^i_{jk} = d_{jk}^i\,.\label{cardy-like}
\end{equation}
We obtain
\begin{equation}
W_{jk} = \sum_{i > N} \frac{d_{jk}^i}{y_j + y_k - y_i}\, \Op_i\,.\quad
(j,k=1,\cdots,N)
\end{equation}
Let ${g'}\,^i\,(i=1,\cdots,N)$ be the choice of parameters for this alternative
convention.  These are related to $g$'s satisfying
(\ref{appendix-linear-beta}) as
\[
{g'}\,^i = g^i + \frac{1}{2} \sum_{j,k=1}^N \frac{d_{jk}^i}{y_j+y_k-y_i}
g^j g^k
\]
to order $g^2$.  (\ref{cardy-like}) is a relation very much like what
Cardy has obtained using UV regularization in
coordinate space \cite{cardy1996scaling}.


\bibliography{paper-short}

\end{document}